# The Rotation-Vibration Spectrum of Diatomic Molecules with the Tietz-Hua Rotating oscillator

M. Hamzavi[1*], A. A. Rajabi[2], K-E Thylwe[3]

[1]Department of Basic Sciences, Shahrood Branch, Islamic Azad University, Shahrood, Iran

[2]Physics Department, Shahrood University of Technology, Shahrood, Iran

[3]KTH-Mechanics, Royal Institute of Technology, S-100 44 Stockholm, Sweden

[*]Corresponding author: Tel.:+98 273 3395270, fax: +98 273 3395270

Email: majid.hamzavi@gmail.com

**Abstract**

The Tietz-Hua (TH) potential is one of the very best analytical model potentials for the vibrational energy of diatomic molecules. By using the Nikiforov-Uvarov (NU) method, we have obtained the exact analytical s-wave solutions of the radial Schrödinger equation (SE) for the TH potential. The energy eigenvalues and the corresponding eigenfunctions are calculated in closed forms. Some numerical results for diatomic molecules are also presented.

## 1. Introduction

The solutions of fundamental dynamical equations are important in many fields of physics and chemistry. It is well known that the exact solutions play an important role in quantum mechanics since they contain all the necessary information regarding the quantum system under consideration. The exact solutions of the SE for a hydrogen atom (Coulombic) and for a harmonic oscillator represent two typical examples in quantum mechanics [1-3]. The Mie-type and pseudoharmonic potentials are also two exactly solvable potentials [4-5]. Many authors exactly solved SE with different potentials and methods [6-17].

The Tietz-Hua (TH) potential is one of the very best analytical model potentials for the vibrational energy of diatomic molecules [18, 19]. The Tietz-Hua potential is much more realistic than the Morse potential in the description of molecular dynamics at moderate and high rotational and vibrational quantum numbers. Also, it may fit the experimental RKR (Rydberg-Klein-Rees) curve more closely than the Morse function does, especially when the potential domain extends to near the dissociation limit [19-22]. Kunc and Gordillo-Vázquez derived analytical expressions for the rotational-vibrational energy levels of diatomic molecules represented by the Tietz-Hua rotating oscillator using the Hamilton-Jacoby theory and the Bohr-Sommerfeld quantization rule [23].

The TH potential is given by

$$V(r) = D \left[ \frac{1 - e^{-b_h (r - r_e)}}{1 - c_h e^{-b_h (r - r_e)}} \right]^2 \qquad (1)$$

with $b_h = \beta(1 - c_h)$, where $r$ is the internuclear distance, $r_e$ is the molecular bond length, $\beta$ is the Morse constant, $D$ is the potential well depth, and $c_h$ is the potential constant [23]. When the potential constant approaches zero, i.e. $c_h = 0$, the TH potential reduces to the Morse potential [24].

This work is arranged as follows: In Section 2, the NU method with all the necessary formulae used in the calculations is briefly introduced and the parametric generalization NU method is displayed in more detail in Appendix A. In Sec. 3 we solve SE and give energy spectra and the corresponding wave functions. Some numerical results are given in this section too. Finally, the relevant conclusions are given in section 4.

## 2. NU method

The NU method can be used to solve second order differential equations with an appropriate coordinate transformation $s = s(r)$ [25]

$$\psi_n''(s) + \frac{\tilde{\tau}(s)}{\sigma(s)} \psi_n'(s) + \frac{\tilde{\sigma}(s)}{\sigma^2(s)} \psi_n(s) = 0, \qquad (2)$$

where $\sigma(s)$ and $\tilde{\sigma}(s)$ are polynomials, at most of second degree, and $\tilde{\tau}(s)$ is a first-degree polynomial. A solution of Eq. (2) is found by a separation of variables, using the transformation $\psi_n(s) = \phi(s) y_n(s)$. It reduces to an equation of the hypergeometric type

$$\sigma(s) y_n''(s) + \tau(s) y_n'(s) + \lambda y_n(s) = 0. \qquad (3)$$

$y_n(s)$ is the hypergeometric-type function whose polynomials solutions are given by Rodrigues relation

$$y_n(s) = \frac{B_n}{\rho(s)} \frac{d^n}{ds^n}\left[\sigma^n(s)\rho(s)\right], \qquad (4)$$

where $B_n$ is the normalization constant and the weight function, $\rho(s)$, must satisfy the condition [25]

$$\frac{d}{ds}w(s) = \frac{\tau(s)}{\sigma(s)}w(s), \qquad w(s) = \sigma(s)\rho(s). \qquad (5)$$

$\phi(s)$ is defined from its logarithmic derivative relation

$$\frac{\phi'(s)}{\phi(s)} = \frac{\pi(s)}{\sigma(s)}. \qquad (6)$$

The function $\pi(s)$ and the parameter $\lambda$, required for this method, are defined as follows

$$\pi(s) = \frac{\sigma' - \tilde{\tau}}{2} \pm \sqrt{\left(\frac{\sigma' - \tilde{\tau}}{2}\right)^2 - \tilde{\sigma} + k\sigma}, \qquad (7a)$$

$$\lambda = k + \pi'(s). \qquad (7b)$$

In order to find the value of $k$, the expression under the square root must be a square of a polynomial. Thus, a new eigenvalue equation is

$$\lambda = \lambda_n = -n\tau' - \frac{n(n-1)}{2}\sigma'', \qquad (8)$$

where

$$\tau(s) = \tilde{\tau}(s) + 2\pi(s), \qquad (9)$$

and its derivative must be negative [25]. In this regard, one can derive the parametric generalization version of the NU method [26] outlined in some detail in Appendix A.

### 3. Solution of radial SE with the TH potential

To study any quantum physical system characterized by the empirical potential given in Eq. (1), we solve the original SE which is given in the well known textbooks [1-2]

$$\left(\frac{P^2}{2\mu} + V(r)\right)\psi(r,\theta,\varphi) = E\psi(r,\theta,\varphi), \qquad (10)$$

where the potential $V(r)$ is taken as a TH potential in Eq. (1), and $\mu$ is the reduced mass. Using the separation method for the wave function $\psi(r,\theta,\varphi) = \frac{R(r)}{r}Y_{lm}(\theta,\varphi)$, we obtain the following radial SE as

$$\left[\frac{d^2}{dr^2}+\frac{2\mu}{\hbar^2}\left(E-D\left[\frac{1-e^{-b_h(r-r_e)}}{1-c_h e^{-b_h(r-r_e)}}\right]^2\right)-\frac{l(l+1)}{r^2}\right]R_{nl}=0. \quad (11)$$

Since the SE with the TH potential has no analytical solution for $l\neq 0$ states, we solve s-wave solutions of Eq. (11), i.e. for $l=0$. Thus Eq. (11) reduces to the following equation

$$\left[\frac{d^2}{dr^2}+\varepsilon-d\left[\frac{1-e^{-b_h(r-r_e)}}{1-c_h e^{-b_h(r-r_e)}}\right]^2\right]R_{n,0}=0, \quad (12)$$

where $\varepsilon=2\mu E/\hbar^2$ and $d=2\mu D/\hbar^2$. To solve Eq. (12) by the NU method, we introduce variables $x=\dfrac{r-r_e}{r_e}$ and $\alpha=b_h r_e$, and use the appropriate transformation with $s=e^{-\alpha x}$. Therefore, Eq. (12) reduces to

$$\frac{d^2 R_{n,0}(s)}{ds^2}+\frac{1-c_h s}{s(1-c_h s)}\frac{dR_{n,0}(s)}{ds}$$
$$+\frac{1}{\alpha^2 s^2(1-c_h s)^2}\left[\varepsilon(1-c_h s)^2-d(1-s)^2\right]R_{n,0}(s)=0 \quad (13)$$

Comparing Eq. (14) and Eq. (A1), we can easily obtain the coefficients $\alpha_i$ ($i=1,2,3$) and the analytical expressions $\xi_j$ ($j=1,2,3$) as follows

$$\begin{aligned}&\alpha_1=1, & &\xi_1=\frac{r_e^2}{\alpha^2}(d-\varepsilon c_h^2)\\ &\alpha_2=c_h, & &\xi_2=\frac{2r_e^2}{\alpha^2}(-\varepsilon c_h+d) \quad (14)\\ &\alpha_3=c_h, & &\xi_3=\frac{r_e^2}{\alpha^2}(d-\varepsilon)\end{aligned}$$

The values of the coefficients $\alpha_i$ ($i=4,5,\ldots,13$) are found from relations (A6-A10, A12, A20-A21, A25-A26) of Appendix A. The specific values of the coefficients $\alpha_i$ ($i=1,2,\ldots,13$) together with $\xi_j$ ($j=1,2,3$) are displayed in table 1. By using (A17), we can obtain the closed form of energy eigenvalues of the s-wave TH potential as

$$(2n+1)\left[\sqrt{\frac{c_h^2}{4}+\frac{2\mu D}{b_h^2\hbar^2}(c_h-1)^2}+c_h\sqrt{\frac{2\mu}{b_h^2\hbar^2}(D-E_{n,0})}\right]+\frac{2\mu D}{b_h^2\hbar^2}(c_h-1)$$
$$+2\sqrt{\left[\frac{c_h^2}{4}+\frac{2\mu D}{b_h^2\hbar^2}(c_h-1)^2\right]\left[\frac{2\mu}{b_h^2\hbar^2}(D-E_{n,0})\right]}+c_h(n^2+3n+0.5)=0 \quad (15)$$

We use the potential parameters from Ref. [23] presented in table 2. Some numerical results of the diatomic molecules $HF$, $N_2$, $I_2$, $H_2$, $O_2$ and $O_2^+$ are given in table 3-5, where they are compared with the Morse potential. Also, we compared the results with Ref. [28]. We use $\hbar c = 1973.29 eV \overset{\circ}{A}$ taken from [4, 11, 27, 28, 29], and the recent conversions: $1cm^{-1} = 1.239841875 eV$ and $1amu = 931.494028 MeV/c^2$ [30]. The authors in Ref. [28] used the following conversions: $1cm^{-1} = 1.23985 eV$ and $1amu = 931.502 MeV/c^2$. As we see from Eq. (1), when $c_h = 0$, the TH potential reduces to the Morse potential and the energy eigenvalues can be obtain as

$$E_{n,0(Morse)} = D\left[1 - \frac{b_h^2 \hbar^2}{8\mu D^2}\left(2n+1-\sqrt{\frac{2\mu D}{b_h^2 \hbar^2}}\right)^2\right] \tag{16}$$

To find the corresponding wave functions, referring to table 1 and relations (A18) and (A22) of appendix A, we find the functions

$$\rho(s) = s^{2\sqrt{\frac{r_e^2}{\alpha^2}(d-\varepsilon)}}(1-s)^{2\left(\sqrt{\frac{1}{4}+\frac{r_e^2 d}{c_h^2 \alpha^2}(c_h-1)^2}-1\right)}, \tag{17}$$

$$\phi(s) = s^{\sqrt{\frac{r_e^2}{\alpha^2}(d-\varepsilon)}}(1-s)^{\frac{1}{2}+\sqrt{\frac{1}{4}+\frac{r_e^2 d}{c_h^2 \alpha^2}(c_h-1)^2}}. \tag{18}$$

Hence, relation (A19) gives

$$y_n(s) = P_n^{\left(2\sqrt{\frac{r_e^2}{\alpha^2}(d-\varepsilon)},\frac{1}{2}+\sqrt{\frac{1}{4}+\frac{r_e^2 d}{c_h^2 \alpha^2}(c_h-1)^2}\right)}(1-2c_h s). \tag{19}$$

By using $R_{n,0}(s) = \phi(s) y_n(s)$, we get the radial wave functions from relation (A24) as

$$R_{n,0}(s) = s^{\sqrt{\frac{r_e^2}{\alpha^2}(d-\varepsilon)}}(1-s)^{\frac{1}{2}+\sqrt{\frac{1}{4}+\frac{r_e^2 d}{c_h^2 \alpha^2}(c_h-1)^2}} P_n^{\left(2\sqrt{\frac{r_e^2}{\alpha^2}(d-\varepsilon)},\frac{1}{2}+\sqrt{\frac{1}{4}+\frac{r_e^2 d}{c_h^2 \alpha^2}(c_h-1)^2}\right)}(1-2c_h s), \tag{20}$$

or, by substituting $s = e^{-\alpha x}$;

$$R_{n,0}(x) = N_{n,0} e^{-\sqrt{r_e^2(d-\varepsilon)}x}(1-e^{-\alpha x})^{\frac{1}{2}+\sqrt{\frac{1}{4}+\frac{r_e^2 d}{c_h^2 \alpha^2}(c_h-1)^2}} P_n^{\left(2\sqrt{\frac{r_e^2}{\alpha^2}(d-\varepsilon)},\frac{1}{2}+\sqrt{\frac{1}{4}+\frac{r_e^2 d}{c_h^2 \alpha^2}(c_h-1)^2}\right)}(1-2c_h e^{-\alpha x}), \tag{21}$$

where $N_{n,0}$ is normalization constant.

## 4. Conclusions

In this article, we have obtained the bound state solutions of the Schrödinger equation for the Tietz-Hua potential by using the parametric generalization of the Nikiforov-Uvarov method. The energy eigenvalues and the corresponding eigenfunctions are obtained by this method. Some numerical results for diatomic molecules are given in tables 3-5 and it is found that when the potential constant $c_h$ goes to zero, the energy levels approach to the Morse potential one.

**References**

[1]. L. I. Schiff: *Quantum Mechanics*, 3rd ed., McGraw-Hill Book Co., New York, 1955.

[2]. R. L. Liboff: *Introductory Quantum Mechanics*, 4th ed., Addison Wesley, San Francisco, CA, 2003.

[3]. M. M. Nieto, AmJ. Phys. **47** (1979) 1067.

[4]. S. M. Ikhdair and R. Sever, J. Mol. Struc.(Theochem) **806** (2007) 155.

[5]. S. M. Ikhdair and R. Sever, J. Mol. Struc.(Theochem) **855** (2008) 13.

[6]. W. C. Qiang and S. H. Dong, Phys. Lett. A **363** (2007) 169.

[7]. S. H. Dong, D.Morales and J. Garcia-Ravelo, Int. J.Mod. Phys. E **16** (2007) 189.

[8]. S. M. Ikhdair and R. Sever, Int. J. Theor. Phys. **46** (2007) 1643.

[9]. S. Ozcelik and M. Simsek, Phys. Lett. A **152** (1991) 145.

[10]. K. J. Oyewumi, Foundations Phys. Lett. **18** (2005) 75.

[11]. O. Bayrak, I. Boztosun and H. Ciftci, Int. J. Quantum Chem. **107** (2007) 540.

[12]. M. R. Setare and E. Karimi, Phys. Scr. **75** (2007) 90.

[13]. S. H. Dong, Phys. Scr. **64** (2001) 273.

[14]. S. H. Dong, Int. J. Theor. Phys. **39** (2000) 1119.

[15]. C. Berkdemir, A. Berkdemir and J. Han, Chem. Phys. Lett. 417 (2006) 326.

[16]. S. M. Ikhdair and R. Sever, Cent. Euro. J. Phys. **5** (2007) 516.

[17]. M. Hamzavi and A. A. Rajabi, to be appear in Int. J. Quant. Chem. (2011).

[18]. T. Tietz, J. Chem. Phys. **38** (1963) 3036.

[19]. W. Hua, Phys. Rev. A **24** (1990) 2524.

[20]. G. A. Natanson, Phys. Rev. A **44** (1991) 3377.

[21]. E. Levin, H. Partridge and J. R. Stallcop, J. Thermophys. Heat Transfer **4** (1990) 469.

[22]. R. T. Pack, J. Chem. Phys. **57** (1972) 4612.

[23]. J. A. Kunc and F. J. Gordillo-Vázquez, J. Phys. Chem. A **101** (1997) 1595.

[24]. P. M. Morse, Phys. Rev.. **34** (1929) 57.

[25]. A. F. Nikiforov and V. B. Uvarov: *Special Functions of Mathematical Physic*, Birkhausr, Berlin, 1988.

[26]. C. Tezcan and R. Sever, Int. J. Theor. Phys. **48** (2009) 337.

[27]. C. Berkdermir and R. Server, J. Math. Chem. **46** (2009) 1122.

[28]. S. M. Ikhdair, Chem. Phys. **361** (2009) 9.

[29]. S. M. Ikhdair and R. Sever, J. Math. Chem. **45** (2009) 1137.

[30]. K. Nakamura et al. (Particle Data Group), J. Phys. G **37** (2010) 075021.

**Appendix A: Parametric Generalization of the Nikiforov-Uvarov method**

The following equation is a general form of the Schrödinger-like equation written for any potential [26]

$$\left[\frac{d^2}{ds^2}+\frac{\alpha_1-\alpha_2 s}{s(1-\alpha_3 s)}\frac{d}{ds}+\frac{-\xi_1 s^2+\xi_2 s-\xi_3}{\left[s(1-\alpha_3 s)\right]^2}\right]\psi_n(s)=0. \quad (A1)$$

We may solve this as follows. When Eq. (A1) is compared with Eq. (2), we get

$$\tilde{\tau}(s)=\alpha_1-\alpha_2 s, \quad (A2)$$

and

$$\sigma(s)=s(1-\alpha_3 s), \quad (A3)$$

also

$$\tilde{\sigma}(s)=-\xi_1 s^2+\xi_2 s-\xi_3, \quad (A4)$$

Substituting these into Eq. (7a), we find

$$\pi(s)=\alpha_4+\alpha_5 s \pm\left[(\alpha_6-k\alpha_3)s^2+(\alpha_7+k)s+\alpha_8\right]^{\frac{1}{2}}, \quad (A5)$$

where

$$\alpha_4=\frac{1}{2}(1-\alpha_1), \quad (A6)$$

$$\alpha_5=\frac{1}{2}(\alpha_2-2\alpha_3), \quad (A7)$$

$$\alpha_6=\alpha_5^2+\xi_1, \quad (A8)$$

$$\alpha_7=2\alpha_4\alpha_5-\xi_2, \quad (A9)$$

$$\alpha_8=\alpha_4^2+\xi_3. \quad (A10)$$

In Eq. (A5), the function under square root must be the square of a polynomial according to the NU method. Thus

$$k_{1,2} = -\left(\alpha_7 + 2\alpha_3\alpha_8\right) \pm 2\sqrt{\alpha_8\alpha_9}, \tag{A11}$$

where, we define

$$\alpha_9 = \alpha_3\alpha_7 + \alpha_3^2\alpha_8 + \alpha_6. \tag{A12}$$

For each $k$, the following $\pi$'s are obtained. For

$$k = -\left(\alpha_7 + 2\alpha_3\alpha_8\right) - 2\sqrt{\alpha_8\alpha_9}, \tag{A13}$$

$\pi$ becomes:

$$\pi(s) = \alpha_4 + \alpha_5 s - \left[\left(\sqrt{\alpha_9} + \alpha_3\sqrt{\alpha_8}\right)s - \sqrt{\alpha_8}\right]. \tag{A14}$$

For the same $k$, from Eqs. (9), (A2) and (A5)

$$\tau(s) = \alpha_1 + 2\alpha_4 - \left(\alpha_2 - 2\alpha_5\right)s - 2\left[\left(\sqrt{\alpha_9} + \alpha_3\sqrt{\alpha_8}\right)s - \sqrt{\alpha_8}\right], \tag{A15}$$

and

$$\tau'(s) = -\left(\alpha_2 - 2\alpha_5\right) - 2\left(\sqrt{\alpha_9} + \alpha_3\sqrt{\alpha_8}\right)$$
$$= -2\alpha_3 - 2\left(\sqrt{\alpha_9} + \alpha_3\sqrt{\alpha_8}\right)\langle 0, \tag{A16}$$

are obtained. When Eq. (A2) is used with Eqs. (A15) and (A16), the following equation is derived:

$$\alpha_2 n - (2n+1)\alpha_5 + (2n+1)\left(\sqrt{\alpha_9} + \alpha_3\sqrt{\alpha_8}\right) + n(n-1)\alpha_3$$
$$+ \alpha_7 + 2\alpha_3\alpha_8 + 2\sqrt{\alpha_8\alpha_9} = 0. \tag{A17}$$

This equation gives the energy spectrum of the desired problem. From Eq. (6)

$$\rho(s) = s^{\alpha_{10}-1}\left(1 - \alpha_3 s\right)^{\frac{\alpha_{11}}{\alpha_3} - \alpha_{10} - 1}, \tag{A18}$$

and consequently, after substitution in Eq. (5),

$$y_n(s) = P_n^{(\alpha_{10}-1,\frac{\alpha_{11}}{\alpha_3}-\alpha_{10}-1)}\left(1 - 2\alpha_3 s\right), \tag{A19}$$

where,

$$\alpha_{10} = \alpha_1 + 2\alpha_4 + 2\sqrt{\alpha_8}, \tag{A20}$$

$$\alpha_{11} = \alpha_2 - 2\alpha_5 + 2\left(\sqrt{\alpha_9} + \alpha_3\sqrt{\alpha_8}\right), \tag{A21}$$

and $P_n^{(\alpha,\beta)}$ are Jacobi polynomials. Using Eq. (4)

$$\phi(s) = s^{\alpha_{12}}\left(1 - \alpha_3 s\right)^{-\alpha_{12}-\frac{\alpha_{13}}{\alpha_3}}, \tag{A22}$$

and the general solution becomes:

$$\psi(s) = \phi(s) y_n(s), \tag{A23}$$

$$\psi(s) = s^{\alpha_{12}}\left(1 - \alpha_3 s\right)^{-\alpha_{12}-\frac{\alpha_{13}}{\alpha_3}} P_n^{(\alpha_{10}-1,\frac{\alpha_{11}}{\alpha_3}-\alpha_{10}-1)}\left(1 - 2\alpha_3 s\right). \tag{A24}$$

Here, alpha parameters are given by:

$$\alpha_{12} = \alpha_4 + \sqrt{\alpha_8}, \tag{A25}$$

and

$$\alpha_{13} = \alpha_5 - \left(\sqrt{\alpha_9} + \alpha_3\sqrt{\alpha_8}\right). \tag{A26}$$

In some problems $\alpha_3 = 0$ [26]. For such problems, when

$$\lim_{\alpha_3 \to 0} P_n^{(\alpha_{10}-1, \frac{\alpha_{11}}{\alpha_3}-\alpha_{10}-1)}(1-\alpha_3)s = L_n^{\alpha_{10}-1}(\alpha_{11}s), \tag{A27}$$

and

$$\lim_{\alpha_3 \to 0}(1-\alpha_3 s)^{-\alpha_{12}-\frac{\alpha_{13}}{\alpha_3}} = e^{\alpha_{13}s}, \tag{A28}$$

the solution given in Eq. (A24) takes the form

$$\psi(s) = s^{\alpha_{12}} e^{\alpha_{13}s} L_n^{\alpha_{10}-1}(\alpha_{11}s). \tag{A29}$$

In some cases, one may need a second solution of Eq. (A1) [26]. In this case, if the same procedure is followed, by using (from Eq. (A11))

$$k = -(\alpha_7 + 2\alpha_3\alpha_8) + 2\sqrt{\alpha_8\alpha_9}, \tag{A30}$$

the solution becomes

$$\psi(s) = s^{\alpha_{12}^*}(1-\alpha_3 s)^{-\alpha_{12}^*-\frac{\alpha_{13}^*}{\alpha_3}} P_n^{(\alpha_{10}^*-1, \frac{\alpha_{11}^*}{\alpha_3}-\alpha_{10}^*-1)}(1-2\alpha_3 s), \tag{A31}$$

and the energy spectrum is obtained from

$$\alpha_2 n - (2n-1)\alpha_5 + (2n+1)\left(\sqrt{\alpha_9} + \alpha_3\sqrt{\alpha_8}\right) + n(n-1)\alpha_3$$
$$+ \alpha_7 + 2\alpha_3\alpha_8 - 2\sqrt{\alpha_8\alpha_9} = 0. \tag{A32}$$

The pre-defined $\alpha$ parameters are:

$$\alpha_{10}^* = \alpha_1 + 2\alpha_4 - 2\sqrt{\alpha_8},$$
$$\alpha_{11}^* = \alpha_2 - 2\alpha_5 + 2\left(\sqrt{\alpha_9} - \alpha_3\sqrt{\alpha_8}\right),$$
$$\alpha_{12}^* = \alpha_4 - \sqrt{\alpha_8},$$
$$\alpha_{13}^* = \alpha_5 - \left(\sqrt{\alpha_9} - \alpha_3\sqrt{\alpha_8}\right). \tag{A33}$$

**Table 1.** The specific values for the parametric constants necessary for the energy eigenvalues and eigenfunctions

| constant | Analytic value |
|---|---|
| $\alpha_1$ | $1$ |
| $\alpha_2$ | $c_h$ |
| $\alpha_3$ | $c_h$ |
| $\alpha_4$ | $0$ |
| $\alpha_5$ | $-\dfrac{c_h}{2}$ |
| $\alpha_6$ | $\dfrac{c_h^2}{4} + \dfrac{r_e^2}{\alpha^2}(d - \varepsilon c_h^2)$ |
| $\alpha_7$ | $\dfrac{2r_e^2}{\alpha^2}(\varepsilon c_h - d)$ |
| $\alpha_8$ | $\dfrac{r_e^2}{\alpha^2}(d - \varepsilon)$ |
| $\alpha_9$ | $\dfrac{c_h^2}{4} + \dfrac{r_e^2 d}{\alpha^2}(c_h - 1)^2$ |
| $\alpha_{10}$ | $1 + 2\sqrt{\dfrac{r_e^2}{\alpha^2}(d - \varepsilon)}$ |
| $\alpha_{11}$ | $2c_h + 2\left(\sqrt{\dfrac{c_h^2}{4} + \dfrac{r_e^2 d}{\alpha^2}(c_h - 1)^2} + c_h\sqrt{\dfrac{r_e^2}{\alpha^2}(d - \varepsilon)}\right)$ |
| $\alpha_{12}$ | $\sqrt{\dfrac{r_e^2}{\alpha^2}(d - \varepsilon)}$ |
| $\alpha_{13}$ | $-\dfrac{c_h}{2} - \left(\sqrt{\dfrac{c_h^2}{4} + \dfrac{r_e^2 d}{\alpha^2}(c_h - 1)^2} + c_h\sqrt{\dfrac{r_e^2}{\alpha^2}(d - \varepsilon)}\right)$ |
| $\xi_1$ | $\dfrac{r_e^2}{\alpha^2}(d - \varepsilon c_h^2)$ |
| $\xi_2$ | $\dfrac{2r_e^2}{\alpha^2}(-\varepsilon c_h + d)$ |
| $\xi_3$ | $\dfrac{r_e^2}{\alpha^2}(d - \varepsilon)$ |

**Table 2.** Model parameters of the diatomic molecules studied in the present work [23].

| Molecule | $c_h$ | $\mu/10^{-23}(g)$ | $b_h(A^{0-1})$ | $r_c(A^0)$ | $\beta(A^{0-1})$ | $D(cm^{-1})$ |
|---|---|---|---|---|---|---|
| $HF$ | 0.127772 | 0.160 | 1.94207 | 0.917 | 2.2266 | 49382 |
| $N_2$ | -0.032325 | 1.171 | 2.78585 | 1.097 | 2.6986 | 79885 |
| $I_2$ | -0.139013 | 10.612 | 2.12343 | 2.666 | 1.8643 | 12547 |
| $H_2$ | 0.170066 | 0.084 | 1.61890 | 0.741 | 1.9506 | 38318 |
| $O_2$ | 0.027262 | 1.377 | 2.59103 | 1.207 | 2.6636 | 42041 |
| $O_2^+$ | -0.019445 | 1.377 | 2.86987 | 1.116 | 2.8151 | 54688 |

**Table 3.** The energy eigenvalues $E_{n,0}$ corresponding to the Tietz-Hua potential for various $n$ for $HF$ and $N_2$ diatomic molecules.

| | $(E_{n,0} - D)(eV)$ | | | | | |
|---|---|---|---|---|---|---|
| | $HF$ | | | $N_2$ | | |
| $n$ | $c_h = 0.127772$ | $c_h = 0$ | $c_h = 0$ [28] | $c_h = -0.032325$ | $c_h = 0$ | $c_h = 0$ [28] |
| 0 | −5.868757846 | −5.868710627 | −5.86875 | -9.7588029855 | −9.758803355 | −9.75887 |
| 5 | −3.660498629 | −3.625603896 | −3.62564 | -8.359551147 | −8.361425901 | −8.36149 |
| 7 | −2.936429314 | −2.878878209 | −2.87892 | -7.829307157 | −7.832693146 | −7.83276 |

**Table 4**. The energy eigenvalues $E_{n,0}$ corresponding to the Tietz-Hua potential for various $n$ for $I_2$ and $H_2$ diatomic molecules.

| n | $(E_{n,0} - D)(eV)$ | | | | | |
|---|---|---|---|---|---|---|
| | $I_2$ | | | $H_2$ | | |
| | $c_h = -0.139013$ | $c_h = 0$ | $c_h = 0$ [28] | $c_h = 0.17006$ | $c_h = 0$ | $c_h = 0$ [28] |
| 0 | −1.542360938 | −1.542360193 | −1.54237 | -4.481571826 | −4.481466458 | -4.47601 |
| 5 | −1.412403230 | −1.412792100 | −1.41280 | -2.281533873 | −2.220195359 | -2.22052 |
| 7 | −1.361840252 | −1.362556277 | −1.36257 | -1.629999641 | −1.535779780 | -1.53744 |

**Table 5.** The energy eigenvalues $E_{n,0}$ corresponding to the Tietz-Hua potential for various $n$ for $O_2$ and $O_2^+$ diatomic molecules.

| | $(E_{n,0} - D)(eV)$ | | | | | |
|---|---|---|---|---|---|---|
| | $O_2$ | | | $O_2^+$ | | |
| $n$ | $c_h = 0.027262$ | $c_h = 0$ | $c_h = 0$ [28] | $c_h = -0.019445$ | $c_h = 0$ | $c_h = 0$ [28] |
| 0 | −5.116333496 | −5.116334833 | −5.11637 | -6.6645687718 | −6.664569808 | −6.66461 |
| 5 | −4.205982010 | −4.204652914 | −4.20469 | -5.559676435 | −5.560725107 | −5.56077 |
| 7 | −3.867392269 | −3.865008092 | −3.86504 | -5.145272749 | −5.147151435 | −5.14720 |